\documentclass[aps,preprint,showpacs,preprintnumbers,amsmath,amssymb,11pt]{revtex4}

\usepackage{makeidx}
\usepackage{graphics}
\usepackage{graphicx,epsfig}
\usepackage{dcolumn}
\usepackage{bm}
\usepackage{rotating}
\usepackage{amsmath}
\usepackage{amssymb}
\usepackage{color}

\begin{document}

\title{An improved helical subgrid-scale model and large-eddy simulation methods in helical turbulence  }

\author{C. P. Yu
}

\affiliation{$^1$Peking University, Beijing 100871, China\\
            }

\date{\today }
\begin{abstract}
For helical isotropic turbulence, an improved two-term helical
subgrid-scale (SGS)
 model is proposed and four types of dynamic methods are
given to do large-eddy simulation (LES), which include the standard
dynamic procedure, the least quatratic sum dynamic procedure, the
dynamic procedure with single constraint of helicity dissipation and
the dynamic one with dual constraints of energy and helicity
dissipation. Tested $a$ $priori$ and $a$ $posteriori$ in both steady
and decaying helical isotropic turbulence, the four types of dynamic
helical models and the dynamic Smogorinsky model are compared with
results of direct numerical simulations (DNS) together. Numerical
results demonstrate that the three new types of dynamic helical
models predict energy and helicity evolution better than the
standard dynamic helical model, and the two constrained helical
models predict the energy and helicity dissipation rates better than
other models. Furthermore, the constrained helical models have
higher correlation to the real SGS stress and have more similar
probability density functions (PDF) to the DNS results. In general,
the two constrained helical models show some more attractive
features than other models.
\end{abstract}

\maketitle

\section{\label{sec1}Introduction}

In turbulent flows, helicity is an important physical quantity. It
is widespread in the motions of the atmosphere, ocean circulation
and other natural phenomena, and also found in leading edge and
trailing vortices shed from wings and slender bodies \cite{1,2,3}.
Helicity, a pseudoscalar quantity, can be defined as
$h=\emph{\textbf{u}}\cdot\bm{\omega}$, where $\emph{\textbf{u}}$ and
$\bm{\omega}$ are the velocity and vorticity of the turbulent flows,
respectively. Similar to the status of energy in the dynamics of
ideal fluids, helicity has the character of  inviscid invariance.
This physical property determines that helicity is an important
quantity in turbulence research.

Recently, researches on helical turbulence have been considerably
forwarded to the fields of theories, experiments and numerical
simulations. Based on the helical decomposition of velocity, the
mechanism of existing a joint forward cascade of energy and helicity
has been explained in theory \cite{4}. Cascades existing in helical
turbulence have space scale and time scale \cite{5,6,7,8}, and the
researches showed that the existing space scale of helicity cascade
was larger than energy cascade. In the inertial range, the joint
cascade of energy and helicity was dominated by the energy cascade
time scale in the low wave number and the helicity cascade time
scale in high wave number. Using direct numerical simulation of
helical isotropic turbulence, energy and helicity flux were studied.
It was shown that helicity flux was more intermittent than the
energy flux and the spatial structure was much finer \cite{9}.

Large-eddy simulation, as an important method, has been widely used
to research turbulent flows. Several kinds of SGS models
\cite{10,11,12,13,14,15,16,17,18} have been proposed so far, such as
eddy-viscosity model, dynamic model, vortex model, \emph{et al}.
Now, among these SGS stress models, the dynamic mixed model is used
most often. Usually, the standard dynamic mixed models can not
predict the energy dissipation properly \cite{19}. Some researches
suggested that it would be much better if some physical constraints
were taken into account \cite{20,21,22}. Specially, a constrained
subgrid-scale stress model was proposed in homogeneous isotropic
turbulence recently \cite{23}, and in view of this constraint, the
numerical results were improved greatly. The constrained condition
of this model fit the physical constraint of energy dissipation.

In helical isotropic turbulence, there exists a joint energy and
helicity cascades. Thinking of the SGS helicity dissipation rate, Y.
Li $et$ $al$. \cite{24} have proposed a two-term helical SGS model
as
\begin{equation}
\tau_{ij}^{mod}=C_1\Delta^2|\widetilde{S}|\widetilde{S}_{ij}+C_2\Delta^3|\widetilde{S}|\widetilde{R}_{ij},\label{eq1}
\end{equation}
where
$\widetilde{S}_{ij}=\frac{1}{2}\left(\partial_j\widetilde{u_i}+\partial_i\widetilde{u_j}\right)$
is the strain-rate tensor at the grid scale $\Delta$, and
$\widetilde{R}_{ij}=\frac{1}{2}\left(\partial_j\widetilde{\omega_i}+\partial_i\widetilde{\omega_j}\right)$
is the symmetric vorticity gradient at $\Delta$. The $C_1$ and $C_2$
are two model coefficients.

In this paper, we propose a rectified two-term helical SGS model
based on Eq. \eqref{eq1}. And at the same time, we also propose
three types of LES methods, a new dynamic procedure (the least
quadratic sum dynamic procedure), the dynamic procedure with single
constraint and the dynamic one with dual constraints. Here, the
three types of new dynamic helical models, the standard dynamic
helical model and the dynamic Smogorinsky model are tested $a$
$priori$ and $a$ $posteriori$. Comparing the results, we can get
some beneficial conclusions.

\section{\label{sec2}Theoretical Analysis for LES model and methods}
\subsection{\label{subsec1}The improved helical SGS model}

In the inertial range of helical turbulence, there exists the
assumption of scale-invariance, and it demands the SGS models to fit
the assumption. To ensure the second term of Eq. \eqref{eq1} scale
invariant in inertial range, we rectify the form of the second term
and the new helical SGS model can be expressed as
\begin{equation}
\tau_{ij}^{mod}=C_1\Delta^2|\widetilde{S}|\widetilde{S}_{ij}+C_2\lambda_{\Delta}^2\Delta|\widetilde{S}|\widetilde{R}_{ij},\label{eq2}
\end{equation}
where $\lambda_{\Delta}^2=15\langle
\widetilde{u_i}\cdot\widetilde{u_i}\rangle/\langle\widetilde{\omega_i}\cdot\widetilde{\omega_i}\rangle
$, and $\langle\cdot\rangle$ denotes an average over directions of
statistical homogeneous field or over pathlines. Similar to the
definition of Taylor microscales, $\lambda_{\Delta}$ can be defined
as Taylor microscales at the scale $\Delta$. In theory, the improved
helical model can predict the energy and helicity dissipation rates
well simultaneously.

To validate the new helical SGS model $a$ $priori$, a DNS of
three-dimensional incompressible homogeneous isotropic helical
turbulence is introduced here. It solves the forced N-S equations
using a pseudo spectral code in a cubic box with periodic boundary
conditions, and the numerical resolution is $512^{3}$. A Guassian
random field is the initial flow condition, and it has an energy
spectrum as
\begin{equation}
E_{0}(k)=A  k^2U_0^2 k_0^{-5}e^{-\frac{2 k^2}{k_0^2}},\label{eq3}
\end{equation}
where $k_{0}=4.5786$  and $U_{0}=0.715$. The whole system is
maintained by a constant energy input rate $\epsilon=0.1$ and a
constant helicity input rate $\eta=0.3$ in the first two wave number
shells. The kinetic viscous $\nu=0.0006$.

In Fig.\ref{fig1}, we present $\lambda_{\delta}^2/\delta$ as a
function of $\delta/\zeta$ for $a$ $priori$, where $\delta$ is the
filter scale varying in the inertial range, and $\zeta$ is the
Kolmogrove scale. From Fig. \ref{fig1}, one can find that the
numerical behavior of $\lambda_{\delta}^2/\delta$ tends
approximately to a constant in the inertial range. Thus, in the
inertial range we have $\lambda_{\Delta}^2\thicksim \Delta$ in the
numerical behavior in Eq. \eqref{eq2}.
\begin{figure}[htbp]
\centering
\epsfig{file=Fig1.eps,bbllx=85pt,bblly=239pt,bburx=700pt,bbury=570pt,width=0.5\textwidth,clip=}
\caption{$\lambda_{\delta}^2/\delta$ distributes with $\delta/\zeta$
for \emph{a priori}. $\zeta$ is the Kolmogrov length
scale.}\label{fig1}
\end{figure}

\subsection{\label{subsec2}The standard dynamic method}

Based on the assumption of scale-invariance in the inertial range,
the standard dynamic models are widely used in large-eddy
simulation. The model coefficients are scale-invariant \cite{25} and
determined by a dynamic procedure which is due to the Germano
identity \cite{14}. The Germano identity can be expressed as
\begin{equation}
L_{ij}=T_{ij}-\overline{\tau}_{ij}=\overline{\widetilde{u}_i\widetilde{u}_j}-\overline{\widetilde{u}}_i\overline{\widetilde{u}}_j,\label{eq4}
\end{equation}
where $L_{ij}$ is the resolved stress,
$\tau_{ij}=\widetilde{u_iu_j}-\widetilde{u}_i\widetilde{u}_j$ is the
SGS stress at the filter scale $\Delta$.
${T}_{ij}=\overline{\widetilde{u_iu_j}}-\overline{\widetilde{u}}_i\overline{\widetilde{u}}_j$
is the SGS stress at the test filter scale $\alpha\Delta$
$(\alpha\neq 1)$. Note that Eq. \eqref{eq2} is the expression of
$\tau_{ij}^{mod}$ , and $T_{ij}^{mod}$ can be written as
\begin{equation}
T_{ij}^{mod}=C_1(\alpha\Delta)^2|\overline{\widetilde{S}}|\overline{\widetilde{S}}_{ij}+C_2\lambda_{\alpha\Delta}^2(\alpha\Delta)|\overline{\widetilde{S}}|\overline{\widetilde{R}}_{ij}.\label{eq5}
\end{equation}
In Eq. \eqref{eq4}, $L_{ij}$, $\tau_{ij}$ and $T_{ij}$ are replaced
by $L_{ij}^{mod}$, $\tau_{ij}^{mod}$ and $T_{ij}^{mod}$, and
substituting $\tau_{ij}^{mod}$ and $T_{ij}^{mod}$ into Eq.
\eqref{eq4}, we can get
\begin{equation}
L_{ij}^{mod}=T_{ij}^{mod}-\overline{\tau}_{ij}^{mod}=C_1M_{ij}+C_2N_{ij},\label{eq6}
\end{equation}
where
\begin{equation}
M_{ij}=\Delta^2\overline{|\widetilde{S}|\widetilde{S}_{ij}}-(\alpha\Delta)^2|\overline{\widetilde{S}}|\overline{\widetilde{S}}_{ij},\label{eq7}
\end{equation}
\begin{equation}
N_{ij}=\lambda_{\Delta}^2\Delta\overline{|\widetilde{S}|\widetilde{R}}_{ij}-\lambda_{\alpha\Delta}^2(\alpha\Delta)|\overline{\widetilde{S}}|\overline{\widetilde{R}}_{ij}.\label{eq8}
\end{equation}
Here, an average square error is introduced as
\begin{equation}
E^{mod}=\langle(L_{ij}-L_{ij}^{mod})^2\rangle.\label{eq9}
\end{equation}
By minimizing Eq. \eqref{eq9}, the model coefficients $C_1$ and
$C_2$ can be obtained.

\subsection{\label{subsec3}The least quadratic sum dynamic method}

In helical turbulence, there exists a joint helicity-energy cascade,
and we need to consider the energy and helicity dissipation
simultaneously. Testing for $a$ $priori$, we find that
$\widetilde{S}_{ij}$ and $\widetilde{R}_{ij}$ have a great
difference in order of magnitude, and thus the energy dissipation
rate $\langle\tau_{ij}\widetilde{S_{ij}}\rangle$ and the helicity
dissipation rate $2\langle\tau_{ij}\widetilde{R_{ij}}\rangle$ also
have a prominent deviation in order of magnitude.

In order to predict energy and helicity evolution more accurately,
we suggest a new type of error as
\begin{equation}
E^{mod}_N=\langle(L_{ij}\overline{\widetilde{S}}_{ij}-L_{ij}^{mod}\overline{\widetilde{S}}_{ij})^2\rangle+\langle(L_{ij}\overline{\widetilde{R}}_{ij}-L_{ij}^{mod}\overline{\widetilde{R}}_{ij})^2\rangle.\label{eq10}
\end{equation}
Minimizing Eq. \eqref{eq10}, we can get the expression of the model
coefficients $C_1$ and $C_2$. The new dynamic procedure is the least
quadratic sum dynamic procedure.

\subsection{\label{subsec4}The constrained method}

In homogeneous isotropic turbulence, the constrained condition on
energy dissipation has been discussed appropriately \cite{23}. In
helical turbulence, the constrained conditions need to be decided by
the energy dissipation and the helicity dissipation jointly. Fig.
\ref{fig2} shows the energy dissipation $\varepsilon_{\delta}$ and
helicity dissipation $\eta_{\delta}$ as a function of
$\delta/\zeta$. We can see that $\varepsilon_{\delta}$ and
$\eta_{\delta}$ are almost constant in the inertial range, and it
fits the assumption of scale-invariance in the inertial range.
\begin{figure}[htbp]
\centering
\epsfig{file=Fig2.eps,bbllx=85pt,bblly=310pt,bburx=700pt,bbury=570pt,width=0.5\textwidth,clip=}
\caption{$\varepsilon_{\delta}$ and $\eta_{\delta}$ distribute with
$\delta/\zeta$ for \emph{a priori}. Line with deltas:
$\varepsilon_{\delta}$; Line with squares: $\eta_{\delta}$
.}\label{fig2}
\end{figure}
In the inertial range of helical isotropic turbulence, the average
energy flux and helicity flux across different scales are almost
invariable and equal to the SGS energy and helicity dissipation
respectively,
\begin{equation}
\varepsilon_{\Delta}=\langle\Pi_E\rangle=-\langle\tau_{ij}\widetilde{S}_{ij}\rangle,\label{eq11}
\end{equation}
and
\begin{equation}
\eta_{\Delta}=\langle\Pi_H\rangle=-2\langle\tau_{ij}\widetilde{R}_{ij}\rangle,\label{eq12}
\end{equation}
where $\Pi_E$ and $\Pi_H$ are the energy and helicity flux through
scale $\Delta$, respectively. Substituting  $\tau_{ij}^{mod}$ for
$\tau_{ij}$ in Eq. \eqref{eq11} and Eq. \eqref{eq12}, we have
\begin{equation}
\varepsilon_{\Delta}=-\langle\tau_{ij}^{mod}\widetilde{S}_{ij}\rangle,\label{eq13}
\end{equation}
and
\begin{equation}
\eta_{\Delta}=-2\langle\tau_{ij}^{mod}\widetilde{R}_{ij}\rangle.\label{eq14}
\end{equation}
From Eq. \eqref{eq4} we can know that the energy and helicity
dissipation rates at scale are
\begin{equation}
\varepsilon_{\alpha\Delta}=-\langle
(L_{ij}+\overline{\tau}_{ij})\overline{\widetilde{S}}_{ij}\rangle,\label{eq15}
\end{equation}
and
\begin{equation}
\eta_{\alpha\Delta}=-2\langle
(L_{ij}+\overline{\tau}_{ij})\overline{\widetilde{R}}_{ij}\rangle.\label{eq16}
\end{equation}
While the model SGS energy and helicity dissipation rates at scale
$\alpha\Delta$ can be expressed as
\begin{equation}
\varepsilon_{\alpha\Delta}=-\langle
T_{ij}^{mod}\overline{\widetilde{S}}_{ij}\rangle,\label{eq17}
\end{equation}
and
\begin{equation}
\eta_{\alpha\Delta}=-2\langle
T_{ij}^{mod}\overline{\widetilde{R}}_{ij}\rangle.\label{eq18}
\end{equation}
Replacing $\overline{\tau}_{ij}$ with $\overline{\tau_{ij}^{mod}}$
in Eq. \eqref{eq15} and Eq. \eqref{eq16}, and then from Eq.
\eqref{eq15}-Eq. \eqref{eq18} we can get the two constraints of
energy and helicity dissipation rates,
\begin{equation}
\langle T_{ij}^{mod}\overline{\widetilde{S}}_{ij}\rangle=\langle
(L_{ij}+\overline{\tau}_{ij})\overline{\widetilde{S}}_{ij}\rangle,\label{eq19}
\end{equation}
and
\begin{equation}
\langle T_{ij}^{mod}\overline{\widetilde{R}}_{ij}\rangle=\langle
(L_{ij}+\overline{\tau}_{ij})\overline{\widetilde{R}}_{ij}\rangle.\label{eq20}
\end{equation}
The derivations of the constraints Eq. \eqref{eq19} and Eq.
\eqref{eq20} are based on the assumption of scale-invariance in the
inertial range, and it demands each term of the helical SGS models
meets the assumption. If it is true, the following relations are
reasonable:
\begin{equation}
\frac{\langle\Delta^2|\widetilde{S}|\widetilde{S}_{ij}\widetilde{S}_{ij}\rangle}{\varepsilon_{\Delta}}=\frac{\langle(\alpha\Delta)^2|\overline{\widetilde{S}}|\overline{\widetilde{S}}_{ij}\overline{\widetilde{S}}_{ij}\rangle}{\varepsilon_{\alpha\Delta}},
\frac{\langle\lambda_{\Delta}^2\Delta|\widetilde{S}|\widetilde{R}_{ij}\widetilde{S}_{ij}\rangle}{\varepsilon_{\Delta}}=\frac{\langle\lambda_{\alpha\Delta}^2(\alpha\Delta)|\overline{\widetilde{S}}|\overline{\widetilde{R}}_{ij}\overline{\widetilde{S}}_{ij}\rangle}{\varepsilon_{\alpha\Delta}},\label{eq21}
\end{equation}
and
\begin{equation}
\frac{\langle\Delta^2|\widetilde{S}|\widetilde{S}_{ij}\widetilde{R}_{ij}\rangle}{\eta_{\Delta}}=\frac{\langle(\alpha\Delta)^2|\overline{\widetilde{S}}|\overline{\widetilde{S}}_{ij}\overline{\widetilde{R}}_{ij}\rangle}{\eta_{\alpha\Delta}},
\frac{\langle\lambda_{\Delta}^2\Delta|\widetilde{S}|\widetilde{R}_{ij}\widetilde{R}_{ij}\rangle}{\eta_{\Delta}}=\frac{\langle\lambda_{\alpha\Delta}^2(\alpha\Delta)|\overline{\widetilde{S}}|\overline{\widetilde{R}}_{ij}\overline{\widetilde{R}}_{ij}\rangle}{\eta_{\alpha\Delta}}.
\label{eq22}
\end{equation}
Now, we introduce four functions $f_1(\delta)$, $f_2(\delta)$,
$h_1(\delta)$ and $h_1(\delta)$, which are
\begin{equation}
f_1(\delta)=\frac{\langle\delta^2|\widetilde{S}|\widetilde{S}_{ij}\widetilde{S}_{ij}\rangle}{\varepsilon_{\delta}},
f_2(\delta)=\frac{\langle\lambda_{\delta}^2\delta|\widetilde{S}|\widetilde{R}_{ij}\widetilde{S}_{ij}\rangle}{\varepsilon_{\delta}},\label{eq23}
\end{equation}
and
\begin{equation}
h_1(\delta)=\frac{\langle\delta^2|\widetilde{S}|\widetilde{S}_{ij}\widetilde{R}_{ij}\rangle}{\eta_{\delta}},
h_2(\delta)=\frac{\langle\lambda_{\delta}^2\delta|\widetilde{S}|\widetilde{R}_{ij}\widetilde{R}_{ij}\rangle}{\eta_{\delta}}.\label{eq24}
\end{equation}

In Fig. \ref{fig3} and Fig. \ref{fig4} we show the distribution of
$f_1(\delta)$, $f_2(\delta)$, $h_1(\delta)$ and $h_1(\delta)$ with
$\delta/\zeta$. We can see the numerical behavior of the four
functions are almost constant in the inertial range, which verifies
the assumption of scale-invariance again and the validity of the
constrained condition Eq. \eqref{eq19} and Eq. \eqref{eq20}.
\begin{figure}[htbp]
\centering
\epsfig{file=Fig3a.eps,bbllx=85pt,bblly=310pt,bburx=700pt,bbury=570pt,width=0.5\textwidth,clip=}
\epsfig{file=Fig3b.eps,bbllx=85pt,bblly=310pt,bburx=700pt,bbury=570pt,width=0.5\textwidth,clip=}
\caption{(a) $f_1(\delta)$, (b) $f_2(\delta)$ distribute with
$\delta/\zeta$ for \emph{a priori}. }\label{fig3}
\end{figure}
\begin{figure}[htbp]
\centering
\epsfig{file=Fig4a.eps,bbllx=85pt,bblly=310pt,bburx=700pt,bbury=570pt,width=0.5\textwidth,clip=}
\epsfig{file=Fig4b.eps,bbllx=85pt,bblly=310pt,bburx=700pt,bbury=570pt,width=0.5\textwidth,clip=}
\caption{(a) $h_1(\delta)$,(b) $h_2(\delta)$ distribute with
$\delta/\zeta$ for \emph{a priori}.}\label{fig4}
\end{figure}

\section{\label{sec3}THE NUMERICAL RESULTS AND ANALYSIS}

In this section, we will give \emph{a} \emph{priori} and \emph{a
posteriori} test of the LES  models, and do some comparison and
analysis. Five SGS models are choosed to compare each other, and
they are dynamic Smogorinsky model (DSM), the standard dynamic
helical model (DHM), the new dynamic helical model (NDSH), the
dynamic helical model with single constraint of helicity dissipation
(CDSH1) and the dynamic helical model with dual constraints of
energy and helicity dissipation (CDSH2).

First of all, we show some results tested \emph{a} \emph{priori}.
\begin{figure}[htbp]
\centering
\epsfig{file=Fig5.eps,bbllx=85pt,bblly=170pt,bburx=700pt,bbury=570pt,width=0.5\textwidth,clip=}
\caption{ The distribution of SGS energy dissipation rate with
$\delta/\zeta$ for \emph{a priori}. Dashed line: CDSH1; dashdotdot
line: CDSH2; line with squares: NDSH; line with deltas: DSH; line
with diamonds: DSM; the bold solid line: DNS. }\label{fig5}
\end{figure}
\begin{figure}[htbp]
\centering
\epsfig{file=Fig6.eps,bbllx=85pt,bblly=170pt,bburx=700pt,bbury=570pt,width=0.5\textwidth,clip=}
\caption{The distribution of SGS helicity dissipation rate with
$\delta/\zeta$ for \emph{a priori}. Dashed line: CDSH1; dashdotdot
line: CDSH2; line with squares: NDSH; line with deltas: DSH; line
with diamonds: DSM; the bold solid line: DNS.}\label{fig6}
\end{figure}
The energy and helicity dissipation rates are calculated \emph{a}
\emph{priori} in Figs. \ref{fig5} and \ref{fig6}, respectively. We
can see the distribution of energy and helicity dissipation rates in
different scales, and in the inertial range the results from CDSH1
and CDSH2 are closer to the DNS result than other models. For energy
dissipation rate, the NDSH also gives a better result than DSH and
DSM. The results from DSH and DSM are almost the same, and we can
draw the conclusion that the second term of DSH has a trivial
contribution to the energy dissipation rate. While in Fig.
\ref{fig6}, we can see the helicity dissipation rates of NDSH and
DSH deviate far from DNS result, which is mainly caused by the
second term of the helical model.

In Fig. \ref{fig7} and Fig. \ref{fig8},
\begin{figure}[htbp]
\centering
\epsfig{file=Fig7.eps,bbllx=85pt,bblly=170pt,bburx=700pt,bbury=570pt,width=0.5\textwidth,clip=}
\caption{The distribution of probability density functions for the
energy flux at the filter scale $\Delta$ for \emph{a priori}. Dashed
line: CDSH1; dashdotdot line: CDSH2; line with squares: NDSH; line
with deltas: DSH; line with diamonds: DSM; the bold solid line:
DNS.}\label{fig7}
\end{figure}
we give the probability density functions (PDF) of the energy flux
$\Pi_E$ and helicity flux $\Pi_H$ at scale $\Delta$ for the five
types of SGS models, respectively. And the PDF from DNS is also
supported here for comparison. Fig. \ref{fig7} reads that the PDFs
of CDSH1 and CDSH2 can predict the backscatters of energy flux,
while other models can not. It is shown in Fig. \ref{fig8} the
similar results to the PDFs of energy flux in Fig. \ref{fig7}, and
only CDSH1 and CDSH2 can capture the backscatters of helicity flux.
\begin{figure}[htbp]
\centering
\epsfig{file=Fig8.eps,bbllx=85pt,bblly=170pt,bburx=700pt,bbury=570pt,width=0.5\textwidth,clip=}
\caption{The distribution of probability density functions for the
helicity flux at the filter scale $\Delta$ for \emph{a priori}.
Dashed line: CDSH1; dashdotdot line: CDSH2; line with squares: NDSH;
line with deltas: DSH; line with diamonds: DSM; the bold solid line:
DNS.}\label{fig8}
\end{figure}

We show in Fig. \ref{fig9} the PDFs of the SGS stress weight
$\tau_{12}$ at scale $\Delta$ for different SGS models \emph{a
priori}. The PDFs from all the models have the similar trend to the
PDF from DNS, and particularly, the PDF from CDSH1 accords well with
that from DNS.
\begin{figure}[htbp]
\centering
\epsfig{file=Fig9.eps,bbllx=85pt,bblly=170pt,bburx=700pt,bbury=570pt,width=0.5\textwidth,clip=}
\caption{The distribution of probability density functions for the
SGS stress weight $\tau_{12}$ for \emph{a priori}. Dashed line:
CDSH1; dashdotdot line: CDSH2; line with square: NDSH; line with
delta: DSH; line with diamond: DSM; the bold solid line:
DNS.}\label{fig9}
\end{figure}

To confirm the validity of these SGS models' applying to LES, we use
these models to perform three dimensional LES of forced and decaying
helical turbulence. It is noting that the resolution of the LES is
$64^3$, and the basic filter length is $3\pi/64$. A Gaussian filter
is also used here. Resolution of the comparing DNS is $512^3$ for
both forced and decaying helical turbulence. The kinetic viscous
$\nu=0.0006$, and the Guassian filter is taken here.

In Figs. \ref{fig10} and \ref{fig11}, we show the steady energy and
helicity spectra from the five types of SGS models and DNS. We can
see from Fig. \ref{fig10} that the energy spectra from the five SGS
models have no remarkable difference, while from Fig.11 we can see
DSH underestimates seriously the helicity spectra close to the grid
scale, and DSM and NDSH a little overestimates the helicity spectra
close to the grid scale. The CDSH1 and CDSH2 predict both energy and
helicity evolution quite well.
\begin{figure}[htbp]
\centering
\epsfig{file=Fig10a.eps,bbllx=85pt,bblly=170pt,bburx=700pt,bbury=570pt,width=0.5\textwidth,clip=}
\epsfig{file=Fig10b.eps,bbllx=85pt,bblly=170pt,bburx=700pt,bbury=570pt,width=0.5\textwidth,clip=}
\caption{Energy spectra. Bold solid line: DNS; (a) the dashed line:
NDSH; line with squares: DSH; line with deltas: DSM (b) the dashed
line: CDSH1; line with squares: CDSH2; line with deltas:
DSM}\label{fig10}
\end{figure}
\begin{figure}[htbp]
\centering
\epsfig{file=Fig11a.eps,bbllx=85pt,bblly=170pt,bburx=700pt,bbury=570pt,width=0.5\textwidth,clip=}
\epsfig{file=Fig11b.eps,bbllx=85pt,bblly=170pt,bburx=700pt,bbury=570pt,width=0.5\textwidth,clip=}
\caption{Helicity spectra. Bold solid line: DNS; (a) the dashed
line: NDSH; line with squares: DSH; line with deltas: DSM (b) the
dashed line: CDSH1; line with squares: CDSH2; line with deltas:
DSM}\label{fig11}
\end{figure}
\begin{figure}[htbp]
\centering
\epsfig{file=Fig12a.eps,bbllx=85pt,bblly=245pt,bburx=700pt,bbury=570pt,width=0.5\textwidth,clip=}
\epsfig{file=Fig12b.eps,bbllx=85pt,bblly=245pt,bburx=700pt,bbury=570pt,width=0.5\textwidth,clip=}
\epsfig{file=Fig12c.eps,bbllx=85pt,bblly=245pt,bburx=700pt,bbury=570pt,width=0.5\textwidth,clip=}
\epsfig{file=Fig12d.eps,bbllx=85pt,bblly=245pt,bburx=700pt,bbury=570pt,width=0.5\textwidth,clip=}
\epsfig{file=Fig12e.eps,bbllx=85pt,bblly=245pt,bburx=700pt,bbury=570pt,width=0.5\textwidth,clip=}
\caption{Energy spectra for decaying helical turbulence ($a$
$posteriori$), at $t=0$, $6\tau_0$, and $12\tau_0$, where $\tau_0$
is the inertial large eddy turnover time scale. Bold line: DNS; the
dashed line: (a) CDSH1, (b) CDSH2, (c) NDSH, (d) DSH, (e) DSM.
}\label{fig12}
\end{figure}

\newpage
\begin{figure}[htbp]
\centering
\epsfig{file=Fig13a.eps,bbllx=85pt,bblly=245pt,bburx=700pt,bbury=570pt,width=0.5\textwidth,clip=}
\epsfig{file=Fig13b.eps,bbllx=85pt,bblly=245pt,bburx=700pt,bbury=570pt,width=0.5\textwidth,clip=}
\epsfig{file=Fig13c.eps,bbllx=85pt,bblly=245pt,bburx=700pt,bbury=570pt,width=0.5\textwidth,clip=}
\epsfig{file=Fig13d.eps,bbllx=85pt,bblly=245pt,bburx=700pt,bbury=570pt,width=0.5\textwidth,clip=}
\epsfig{file=Fig13e.eps,bbllx=85pt,bblly=245pt,bburx=700pt,bbury=570pt,width=0.5\textwidth,clip=}
\caption{Helicity spectrafor decaying helical turbulence ($a$
$posteriori$), at $t=0$, $6\tau_0$, and $12\tau_0$. Bold line: DNS;
the dashed line: (a) CDSH1, (b) CDSH2, (c) NDSH, (d) DSH, (e)
DSM.}\label{fig13}
\end{figure}

Fig. \ref{fig12} and Fig. \ref{fig13} show the time evolutions of
the decaying energy and helicity spectra of the SGS models, and they
all start from the same fully statistical steady state. In Fig.
\ref{fig12}, we can see that CDSH1 underestimates the energy spectra
greatly close to the grid scale, and the energy spectra of other
models have trivial difference. Helicity is not always positive,
which are caused by the helicity's property of pseudoscalar, and
thus it is shown in Fig. \ref{fig13} that the helicity spectra
display the character of fluctuations. Also similar to the steady
case, the DSH underestimates the helicity close the grid scale, and
the results from other models have no obvious difference.

\newpage
\section{\label{sec4}CONCLUSIONS}

In this paper, we firstly improve the helical SGS model based on an
existing helical model to ensure the helical models scale-invariant
in the inertial range. Then, we propose a new dynamic method and two
types of constrained dynamic methods to predict the energy and
helicity dissipation well simultaneously. Using the improved helical
SGS model with the three new dynamic methods and the standard
dynamic mehod, we have proposed four types of dynamic helical models
and compared them with DSM model and DNS.

Through testing for $a$ $priori$ and $a$ $posteriori$, we have found
that the constrained dynamic helical models predict energy and
helicity dissipation rates well and have high correlation with the
real SGS stress. At the same time, CDSH1 and CDSH2 can predict the
energy and helicity backscatter. NDSH also has some improvement
contrasting with DSH, such as predicting the helicity evolution and
energy dissipation rate.

In short, the constrained dynamic helical models are quite fit to
use in the large eddy simulation of helical isotropic turbulence,
and also fit to apply into other systems, such as rotational
turbulence and magnetohydrodynamics, $et$ $al$.

\section{\label{sec5}acknowledgement}
Here, we give thanks to Shiyi Chen, Zuoli Xiao, Yipeng Shi and
Wanhai Liu for the beneficial discussions.

\end{document}